\documentclass[%
 reprint,
 amsmath,amssymb,
pra,
]{revtex4-1}
\usepackage{color}
\usepackage{graphicx}
\usepackage{dcolumn}
\usepackage{bm}
\usepackage{hyperref}
\usepackage{natbib}
\usepackage{epstopdf}

\begin{document}

\preprint{APS/123-QED}

\title{Unconventional photon blockade in three-mode optomechanics}

\author{Bijita Sarma}
 \email{s.bijita@iitg.ernet.in}
\author{Amarendra K. Sarma}%
 \email{aksarma@iitg.ernet.in}
\affiliation{%
 Department of Physics, Indian Institute of Technology Guwahati, Guwahati-781039, Assam, India}


%

\date{\today}

\begin{abstract}
We analyze the photon correlations in an optomechanical system containing two nonlinear optical modes and one mechanical mode which are coupled via a three-mode mixing. Under a weak driving condition, we determine the optimal conditions for photon antibunching in the weak Kerr-nonlinear regime and we find that the analytical calculations are consistent with the numerical results. The photon blockade effect is attributed to destructive quantum interference in the two-photon excitation pathways created as a result of the three-mode interaction. 

\begin{description}
\item[PACS numbers] 42.50.Ar, 42.65.-k,42.50.Lc,07.10.Cm
\verb||
\end{description}
\end{abstract}

\pacs{Valid PACS appear here}
\maketitle


\section{\label{sec:level1}INTRODUCTION}
The field of cavity optomechanics research, that deals with the effects resulting from light-matter interaction among cavity modes and mechanical motion, has attracted considerable attention in recent years \cite{aspelmeyer2014cavity}. This is primarily owing to its possible futuristic applications in quantum sensors, quantum information processing, solid-state implementation of quantum memory and so on, apart from being a viable tool for testing fundamentals of quantum mechanics \cite{aspelmeyer2012quantum, meystre2013short}. 
Following the early theories on cavity optomechanical cooling of mechanical resonators, recent progress in optomechanical experiments has enabled the realization of mechanical resonators near to the ground state \cite{metzger2004cavity, gigan2006self, arcizet2006radiation, kleckner2006sub, corbitt2007optical, schliesser2008resolved, thompson2008strong, wilson2009cavity, o2010quantum, chan2011laser, sarma2016ground}. This has opened up newer avenues for quantum applications of optomechanical systems \cite{schmidt2012optomechanical, stannigel2010optomechanical, safavi2011proposal}. Recently, cavity optomechanical systems have been studied for its inherent nonlinear coupling to achieve photon blockade \cite{rabl2011photon, nunnenkamp2011single}.

Photon blockade arises from the anharmonicity in energy eigenvalues of an optical mode, which can be introduced via nonlinear interactions. Due to the anharmonicity, resonant excitation of one photon prohibits other photons from simultaneous excitation, giving rise to sub-Poissonian light. The early theories and experiments on photon blockade dealt with atom-coupled cavities \cite{birnbaum2005photon, dayan2008photon}, or quantum dot-coupled cavity QED systems \cite{faraon2008coherent}, or cavities with Kerr-type nonlinearity \cite{imamoglu1997strongly}. After that, there have been several studies on photon blockade in optical waveguides \cite{chang2008crystallization}, coupled cavities \cite{hartmann2006strongly, greentree2006quantum, angelakis2007photon}, qubit-cavity systems \cite{miranowicz2014state}, circuit-QED \cite{lang2011observation, hoffman2011dispersive, liu2014blockade}, gain cavity \cite{zhou2018zero}, and multiphoton blockade in some systems \cite{miranowicz2013two, hovsepyan2014multiphoton, deng2015enhancement}. Numerous possible quantum device designs such as: single-photon transistors \cite{hong2008single}, quantum repeaters \cite{han2010quantum}, quantum gates \cite{wu2010implementation}, quantum-optical Josephson interferometer \cite{gerace2009quantum}, fermionization of photons \cite{carusotto2009fermionized}, and crystallization of polaritons \cite{hartmann2010polariton} rely on the phenomenon of photon blockade. In fact, generation of single photons plays a central role in light-based quantum computation and cryptography \cite{knill2001scheme,duan2001long,kimble2008quantum,o2009photonic}.

Photon blockade in an optomechanical system was studied recently, where due to the photon-phonon nonlinear interaction, realization of antibunched sub-Poissonian light was predicted \cite{rabl2011photon}. Subsequently photon blockade \cite{komar2013single, liao2013photon, wang2015tunable}, as well as phonon blockade \cite{liu2010qubit, didier2011detecting, miranowicz2016tunable} have been studied in various optomechanical and nanomechanical systems. However, similar to cavity-QED systems, realization of optomechanical photon blockade demands the criterion of strong-coupling, where the single-photon optomechanical coupling is strong enough to overcome system losses, in order to produce sufficient anharmonicity in the energy-levels \cite{rabl2011photon}. Reaching this strong-coupling regime is a long-sought-after goal in cavity optomechanics, however only with a few realizations like cold-atomic clouds in optomechanical cavity \cite{brennecke2008cavity, gupta2007cavity}, where this requirement has been met till date.


Recently, another mechanism for photon blockade which does not require the strong-coupling condition to hold, is invoked by Liew and Savona in coupled-polaritonic systems \cite{liew2010single}. This method, named as the unconventional photon blockade, is based on quantum interference, in which strong photon antibunching was predicted with nonlinearity much smaller than the decay rate of the photonic modes \cite{bamba2011origin}. Afterwards, it has been studied in other systems including coupled nonlinear photonic molecules \cite{bamba2011origin}, coupled cavities with Kerr-type nonlinearity \cite{shen2015exact, ferretti2013optimal, flayac2015all, shen2015tunable, flayac2016single, zou2018photon}, coupled optomechanical cavities \cite{xu2013antibunching, savona2013unconventional}, coupled quantum dot-cavity system \cite{tang2015quantum}, bimodal cavity \cite{majumdar2012loss, zhang2014optimal}, weakly nonlinear photonic molecules \cite{xu2014strong}, Gaussian squeezed states \cite{lemonde2014antibunching}, and with second-order nonlinearity \cite{gerace2014unconventional, zhou2015unconventional, sarma2017quantum}.
Recently, unconventional photon blockade has also been realized experimentally in a quantum dot cavity system \cite{snijders2018single}.

In this paper, we study photon correlations in a nonlinear optomechanical cavity containing two optical modes and one mechanical mode which are cross-coupled by a three-mode interaction. We show that even when the Kerr-nonlinearities in the optical modes are weak, due to the three-mode interaction, distinct two-photon excitation pathways arise which gives rise to strong photon antibunching via unconventional photon blockade.
The remainder of the paper is organized as follows. In Sec. II, we describe the model and derive the optimal conditions for photon blockade. In Sec. III, we calculate the equal-time second-order correlation function as well as the two-time correlation function for the higher-frequency cavity mode and analyze the photon blockade characteristics. We also discuss the effects of temperature and pure-dephasing induced decoherences. The results are summarized in Sec. IV.

\section{\label{sec:level1} Model and theory}
\begin{figure}[!]
	\centering
	\includegraphics [trim={0cm 0cm 0cm 0cm},width =0.9\linewidth]{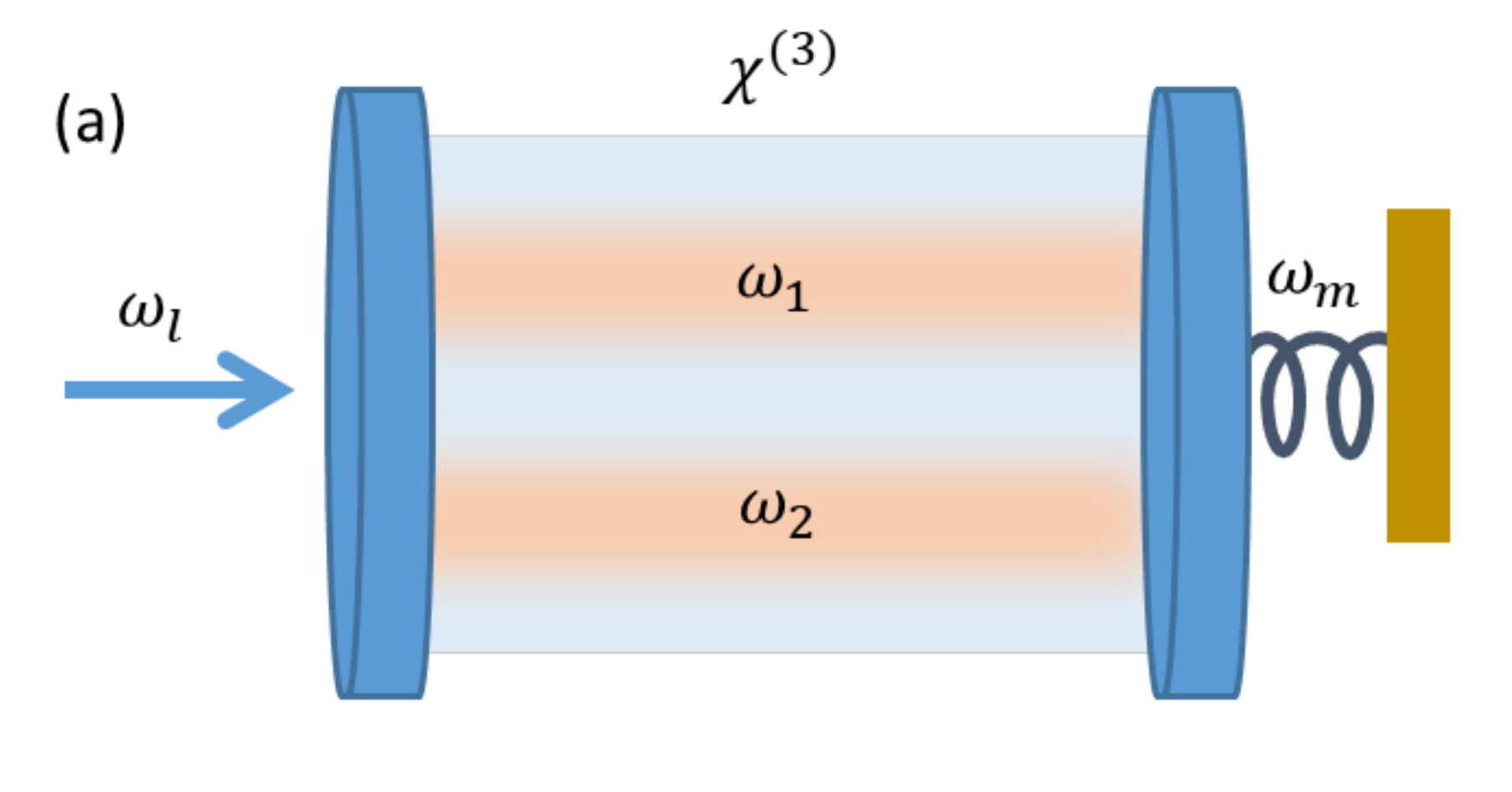}
	\includegraphics [trim={0cm 0cm 0cm 0cm},width =0.8\linewidth]{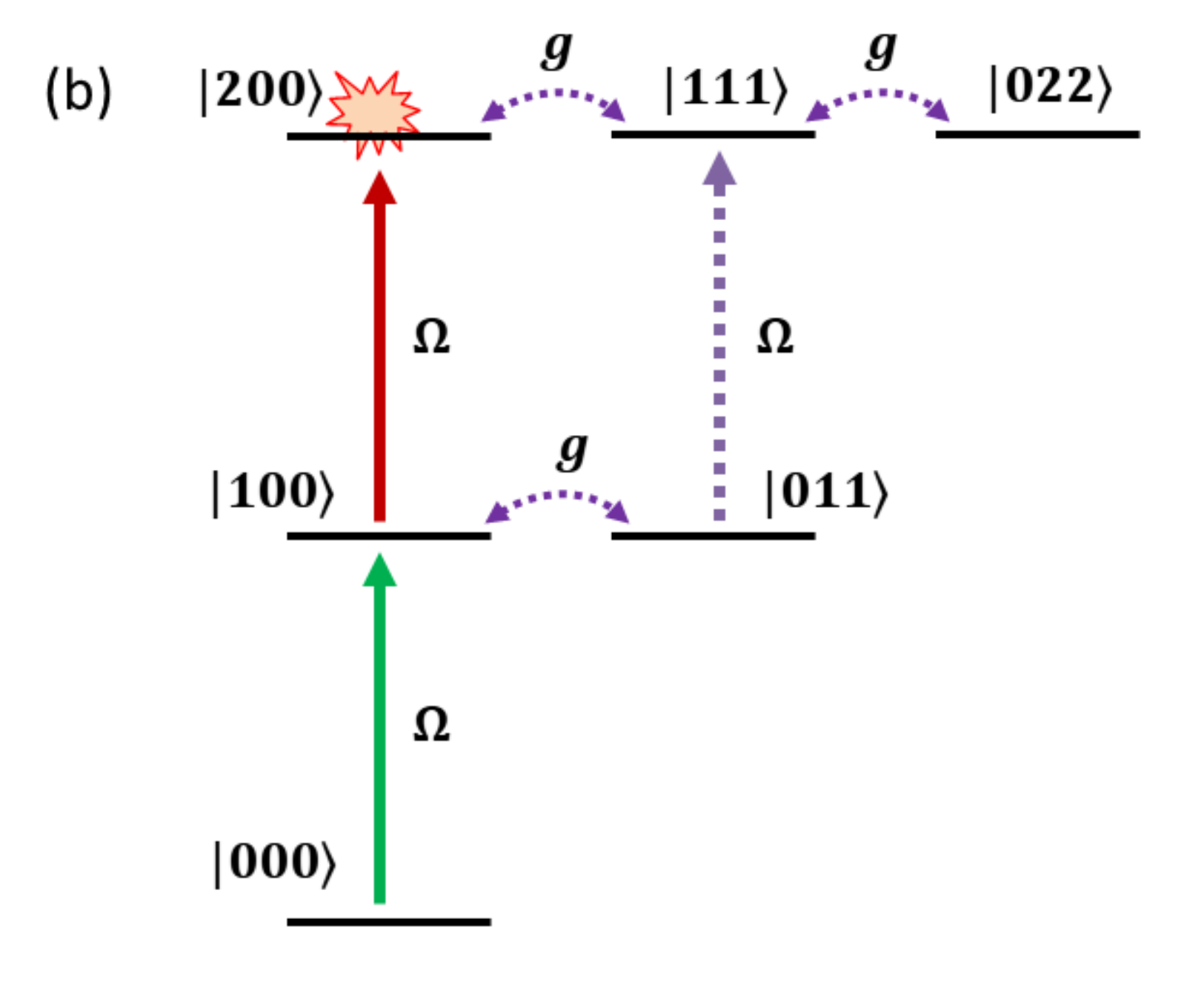}
	\caption {(Color online) (a) Schematic diagram of the optomechanical cavity with two optical modes and a single mechanical mode, (b) The low energy-levels of the system for a weak drive and low temperature.}
	\label {upb1:fig1}
\end{figure}

We consider a nonlinear optomechanical system as shown in Fig.~\ref{upb1:fig1}(a), that contains two cavity modes with frequencies, $\omega_1$ and $\omega_2$, and one mechanical mode with frequency $\omega_m$. The Hamiltonian of the system reads 
\begin{align} \label{upb1:eq1}
\nonumber
H_0\ =& \ \omega_1 a_1^\dagger a_1+ \omega_2 a_2^\dagger a_2+ \omega_m b^\dagger b + U a_1^\dagger a_1^\dagger a_1 a_1 \\
\nonumber
& + U a_2^\dagger a_2^\dagger a_2 a_2
+ g(a_1^\dagger a_2 + a_2^\dagger a_1)(b+b^\dagger)\\
& + \Omega_1(a_1^\dagger e^{-i\omega_l t}+a_1 e^{i\omega_l t}) +
 \Omega_2(a_2^\dagger e^{-i\omega_l t} + a_2 e^{i\omega_l t}),
\end{align}
where $a_1$ ($a_1^\dagger$), $a_2$ ($a_2^\dagger$), and $b$ ($b^\dagger$) are the annihilation (creation) operators for the two cavity modes with decay rates $\kappa_1$ and $\kappa_2$, and the mechanical mode with damping rate $\gamma$, respectively. Here, $U$ is the strength of the Kerr-nonlinearity experienced by both the optical modes. We assume the difference between the two cavity mode frequencies to be equal to the mechanical frequency, i.e.~$\omega_1 - \omega_2 = \omega_m$, so that the cavity modes can be cross-coupled by the optomechanical interaction \cite{chang2011slowing}. The coupling is characterized by the rate, $g$, and is also proportional to the mechanical displacement. The last two terms in Eq.~\eqref{upb1:eq1} describe the driving input fields and its interaction with the two cavity modes. For simplicity, we will assume that $\kappa_1 = \kappa_2 = \kappa$, for the rest of the paper. 

In a rotating frame at the laser frequency, $\omega_l$, the Hamiltonian is transformed to
\begin{align} \label{upb1:eq2}
\nonumber
H \ =& \ \Delta a_1^\dagger a_1 + (\Delta - \omega_m) a_2^\dagger a_2 + \omega_m b^\dagger b + U a_1^\dagger a_1^\dagger a_1 a_1 \\
\nonumber
& + U a_2^\dagger a_2^\dagger a_2 a_2
+ g(a_1^\dagger a_2 + a_2^\dagger a_1)(b + b^\dagger) \\
& + \Omega_1(a_1^\dagger +a_1) +
 \Omega_2(a_2^\dagger +a_2),
\end{align}
where, $\Delta = \omega_1 - \omega_l$, is the detuning of the cavity mode $a_1$ from the laser drive. Now, we transform the Hamiltonian to a frame defined by the unitary transformation, $U= \exp[-i\omega_m t(b^\dagger b - a_2^\dagger a_2)]$. 
Assuming that the coupling rate is much lower in comparison to the mechanical resonator frequency, i.e.~$\omega_m\gg g$, under a rotating-wave approximation, the transformed Hamiltonian is obtained as
\begin{align} \label{upb1:eq3}
\nonumber
H\ =& \ \Delta (a_1^\dagger a_1+ a_2^\dagger a_2) + U a_1^\dagger a_1^\dagger a_1 a_1 + U a_2^\dagger a_2^\dagger a_2 a_2 \\
& + g(a_1^\dagger a_2 b+ a_1 a_2^\dagger b^\dagger)+
 \Omega_1(a_1^\dagger +a_1).
\end{align}
This Hamiltonian indicates a three-mode interaction among the two optical modes and the mechanical mode, in which one photon from the mode, $a_1$, is annihilated to create one photon in the mode, $a_2$, and one phonon in the mechanical mode, $b$. In the reverse process, one photon from the mode, $a_2$, and one phonon in the mode, $b$ of the mechanical resonator are annihilated to create one photon in the mode, $a_1$. In the following, we intend to study the photon antibunching effect in the mode, $a_1$, arising as a result of this three-mode mixing Hamiltonian.

Photon antibunching would be studied by analyzing the normalized zero-time-delay second-order correlation function, given by
\begin{align} \label{upb1:eq4} 
	g_a^{(2)}(0)= \frac{\langle a_1^\dagger (t) a_1^\dagger (t) a_1(t) a_1(t)\rangle}{\langle a_1^\dagger (t) a_1(t)\rangle^2}.
\end{align}	
This quantity characterizes the joint probability of detecting two photons at the same time, which can be calculated numerically from the Lindblad master equation. The master equation for the driven-dissipative system is given by
\begin{align} \label{upb1:eq5}	
	\dot{\rho}=i[\rho, H] + L_1(\rho) + L_2(\rho) + L_b(\rho),
\end{align}
where, $L_1(\rho)= \frac{\kappa}{2}(2a_1 \rho a_1^\dagger - a_1^\dagger a_1 \rho -\rho a_1^\dagger a_1)$, $L_2(\rho)= \frac{\kappa}{2}(2a_2 \rho a_2^\dagger - a_2^\dagger a_2 \rho -\rho a_2^\dagger a_2)$, and $L_b(\rho)= \frac{\kappa}{2} (n_\textrm{th} +1) (2b \rho b^\dagger - b^\dagger b \rho -\rho b^\dagger b) + \frac{\kappa}{2} n_\textrm{th} (2b^\dagger \rho b - b b^\dagger \rho -\rho b b^\dagger)$, are the Liouvillian operators for the two optical modes and the mechanical mode respectively. Here, $n_\textrm{th} = 1/[\exp(\hbar \omega_m/k_B T)-1]$ denotes the thermal phonon number in the mechanical mode at the bath temperature, $T$. The steady-state value of $g_a^{(2)}(0)$ can be found numerically by solving the master equation and then from the steady state density matrix operator as, $g_a^{(2)}(0)= \textrm{Tr} (\rho a_1^\dagger a_1^\dagger a_1 a_1)/[\textrm{Tr} (\rho a_1^\dagger a_1)]^2$.

In addition to the master equation approach, optimal conditions for photon blockade can be determined in the following manner. 
When the driving field is very weak in comparison
to the Kerr nonlinearity, and the temperature is also very low,
then, only the lower energy levels of
the cavity and the mechanical modes are occupied \cite{wang2015tunable}, as shown in Fig.~\ref{upb1:fig1}(b).
Considering the allowed low-energy transitions given by the Hamiltonian in Eq.~\eqref{upb1:eq3}, the truncated state of the system is given by \cite{bamba2011origin}
\begin{align} \label{upb1:eq6}
\nonumber	
|\psi\rangle= C_{000}|000\rangle+C_{100}|100\rangle+C_{011}|011\rangle\\+C_{200}|200\rangle+C_{111}|111\rangle+C_{022}|022\rangle,
\end{align}
where, $C_{a_1a_2b}$'s are the amplitudes of the quantum states for which the corresponding occupation probability is given by $|C_{a_1a_2b}|^2$. The values of the coefficients can be determined by solving the Schr\"{o}dinger equation, $i\frac{d|\psi\rangle}{dt}=H'|\psi\rangle$, where $H'$ is the non-Hermitian Hamiltonian containing the optical decay and mechanical damping terms
\begin{align} \label{upb1:eq7}
\nonumber
H'\ =& \ (\Delta -i\frac{\kappa}{2})(a_1^\dagger a_1+ a_2^\dagger a_2)-i\frac{\gamma}{2}b^\dagger b + U a_1^\dagger a_1^\dagger a_1 a_1\\
&  + U a_2^\dagger a_2^\dagger a_2 a_2 + g(a_1^\dagger a_2 b+ a_1 a_2^\dagger b^\dagger)+
 \Omega_1(a_1^\dagger +a_1).
\end{align}
For a weak drive, a set of equations for the coefficients is obtained from the Schr\"{o}dinger equation 
\begin{align} \label{upb1:eq8}
\nonumber	
i\frac{\partial C_{100}}{\partial t}\ =& \ \left(\Delta-i\frac{\kappa}{2}\right) C_{100}+gC_{011}+\Omega(C_{000}+\sqrt{2}C_{200}),\\
\nonumber
i\frac{\partial C_{011}}{\partial t}\ =& \ \left(\Delta-i\frac{\kappa+\gamma}{2}\right) C_{011}+gC_{100}+\Omega C_{111},\\
\nonumber
i\frac{\partial C_{200}}{\partial t}\ =& \ 2\left(\Delta + U -i\frac{\kappa}{2}\right) C_{200}+\sqrt{2}gC_{111}+\sqrt{2}\Omega C_{100},\\
\nonumber
i\frac{\partial C_{111}}{\partial t}\ =& \ \left[2\left(\Delta-i\frac{\kappa}{2}\right)-i\frac{\gamma}{2}\right] C_{111}+g(2 C_{022} + \sqrt{2}C_{200})\\
\nonumber
& +\Omega C_{011},\\
i\frac{\partial C_{022}}{\partial t}\ =& \ 2\left(\Delta + U - i\frac{\kappa+\gamma}{2}\right) C_{022} + 2gC_{111}.
\end{align}
Under the weak driving assumption, one can consider that $\{C_{111}, C_{022}, C_{200}\} \ll \{C_{100}, C_{011}\} \ll C_{000}$. Now, considering $\gamma \ll \kappa$ for typical optomechanical systems and substituting $C_{200} = 0$, the optimal parameters for complete photon antibunching in the mode $a_1$ is given by (see Appendix):
\begin{align} \label{upb1:eq9}
\nonumber
\Delta_\textrm{opt}\ =& \ \pm \frac{1}{2} \sqrt{2 \sqrt{g^2 (5 g^2 + 2 \kappa^2)} - 4 g^2 - \kappa^2},\\
U_\textrm{opt}\ =& \ \frac{\Delta(4 \Delta^2 + 2 g^2 +5 \kappa^2)}{2(2 g^2 - \kappa^2)}.
\end{align}	
These optimal conditions correspond to the situation, where different transition paths leading to two-photon excitation in the mode $a_1$ interferes destructively, as shown in Fig.~\ref{upb1:fig1}(b).

\begin{figure}
	\centering
	\includegraphics[width=\linewidth]{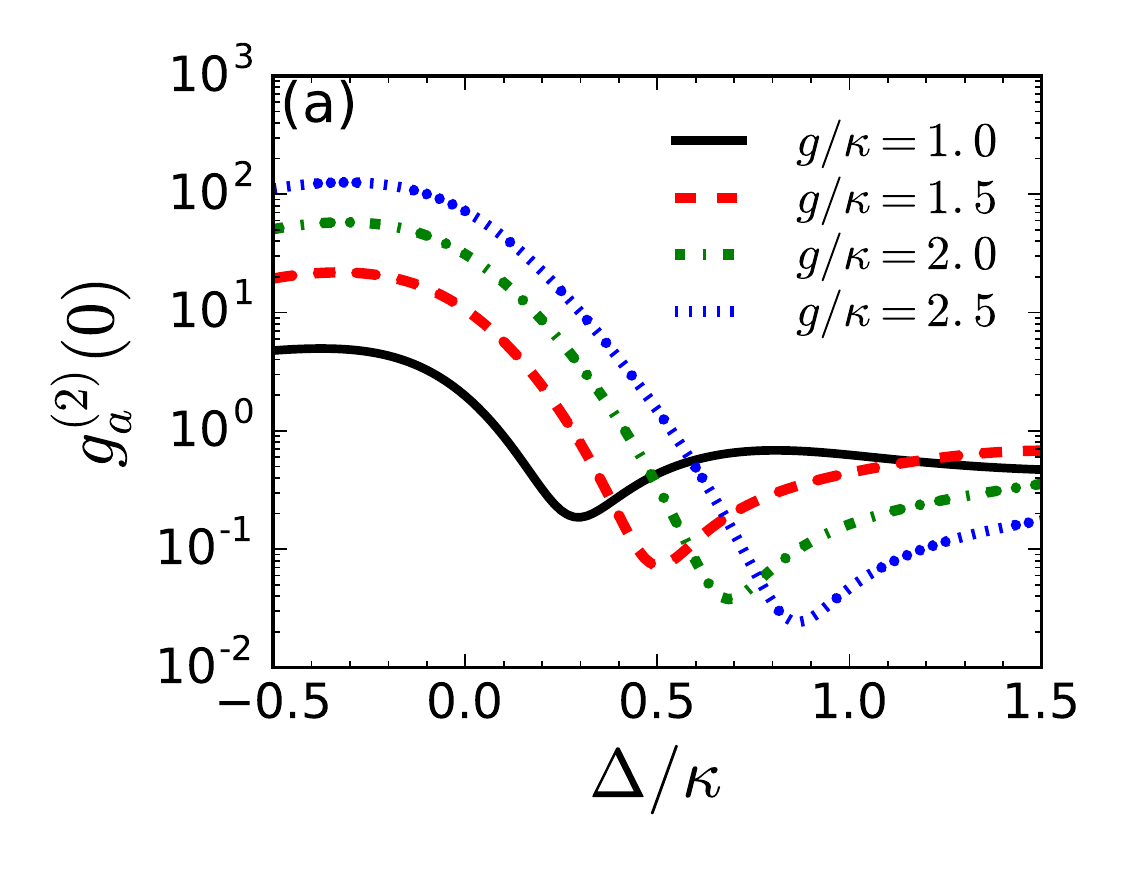}
	\includegraphics[width=\linewidth]{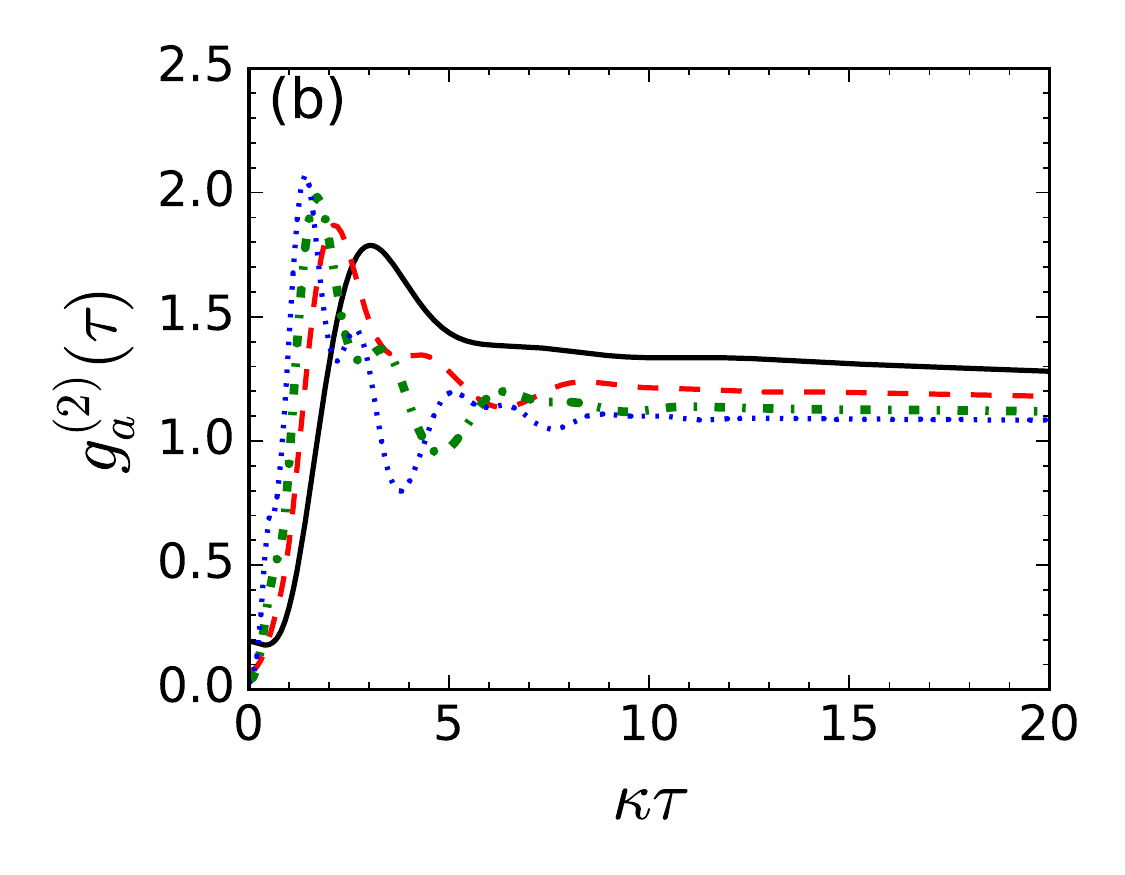}
	\caption{(Color online) Plot of second-order correlation function, $g_a^{(2)}(0)$ at $T = 0$ as a function of normalized detuning $\Delta/\kappa$ for different values of $g/\kappa$. The nonlinearity is considered to be $U_\textrm{opt}/\kappa = 0.98$, $0.71$, $0.69$, and $0.74$ for the normalized values of coupling, $g/\kappa = 1$, $1.5$, $2$ and $2.5$ respectively. (b) Plot of two-time correlation function, $g_a^{(2)}(\tau)$. The values of $U$ is considered same as in (a). The nonlinear strength is considered as $\Delta_\textrm{opt}/\kappa = 0.27$, $0.47$, $0.66$, and $0.84$ respectively.}
	\label{upb1:fig2}
\end{figure}
\section{\label{sec:level1} Results}
\begin{figure*}[!]
	\centering
	\includegraphics[width=0.45\linewidth]{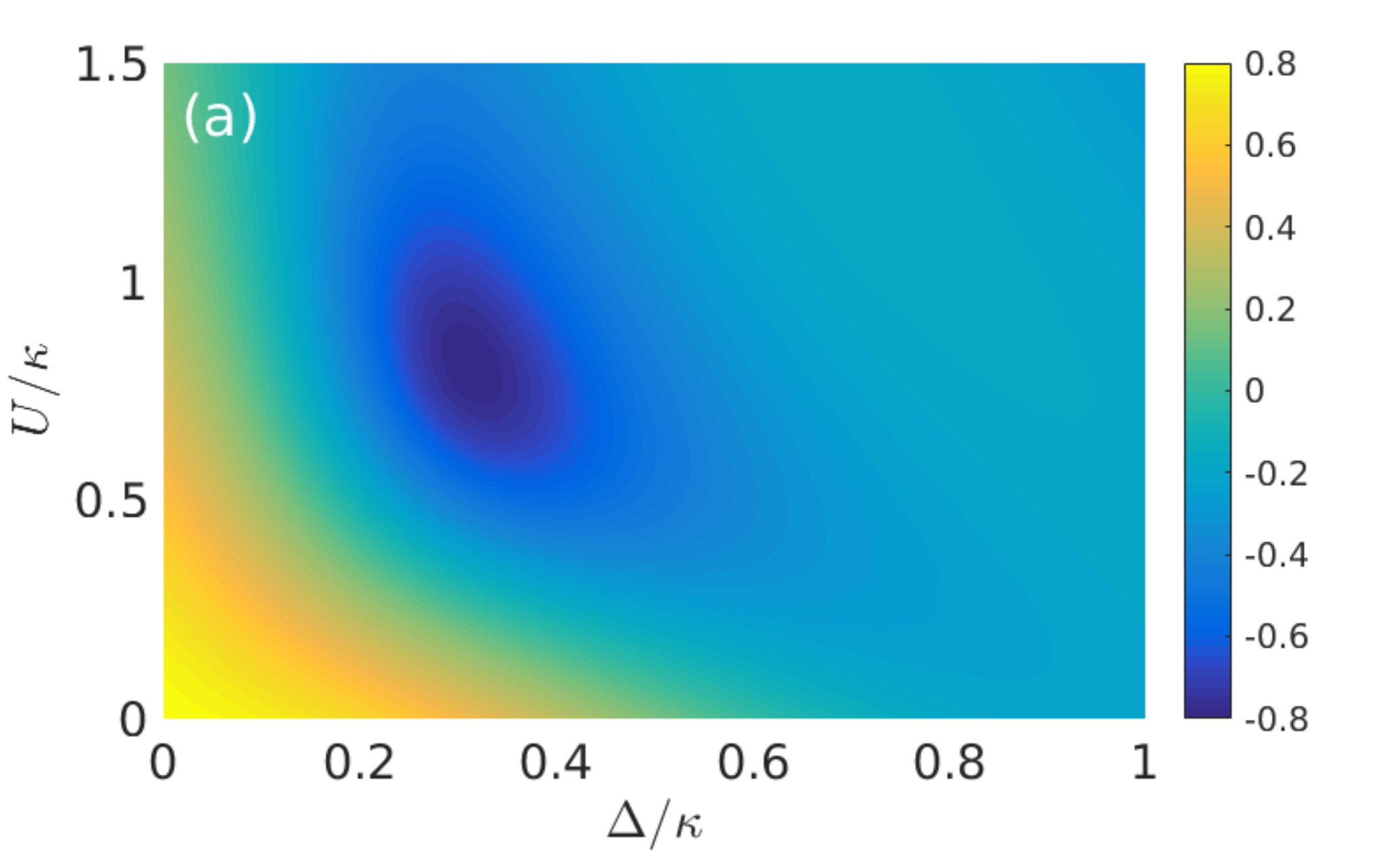}
	\includegraphics[width=0.45\linewidth]{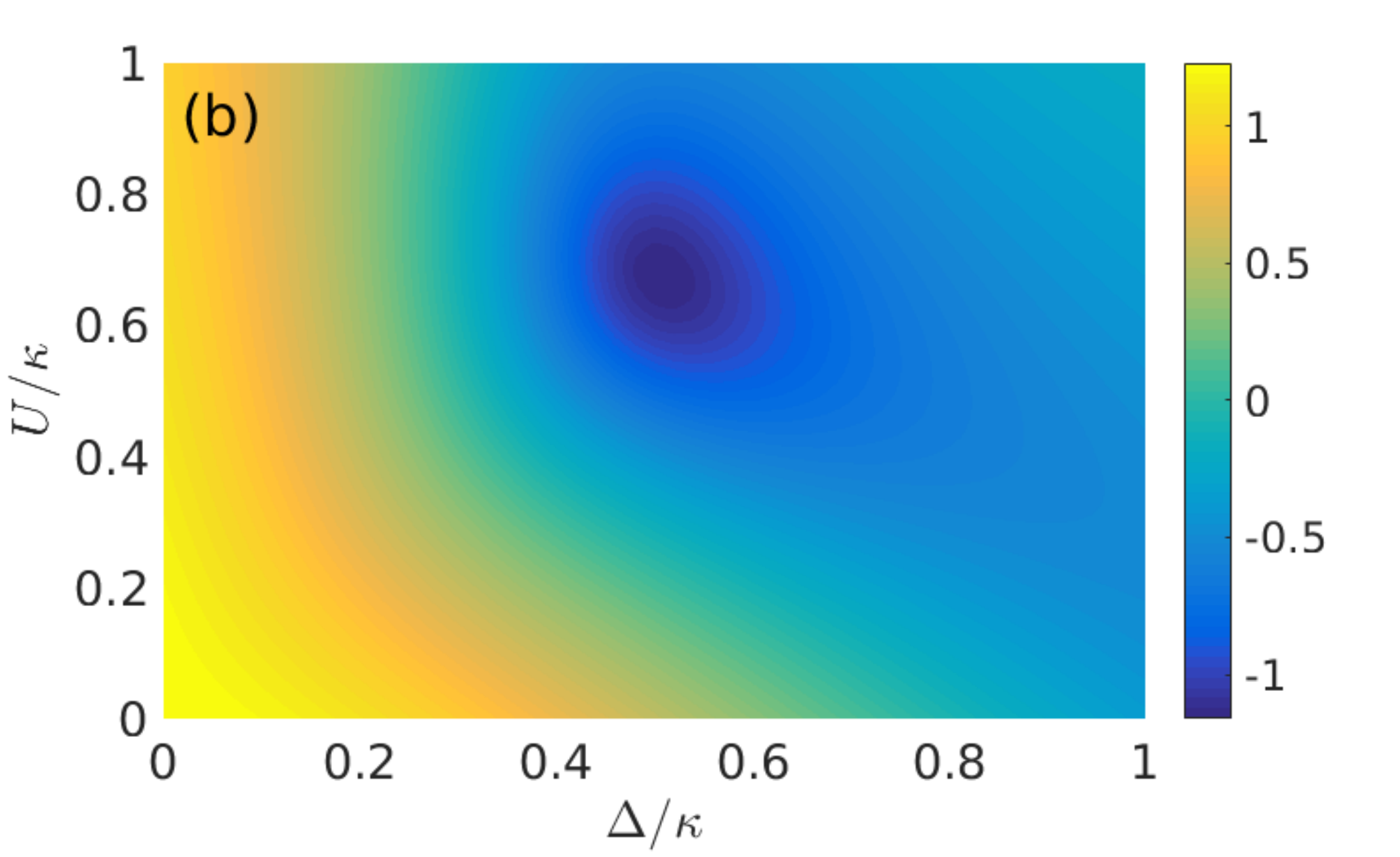}
	\includegraphics[width=0.45\linewidth]{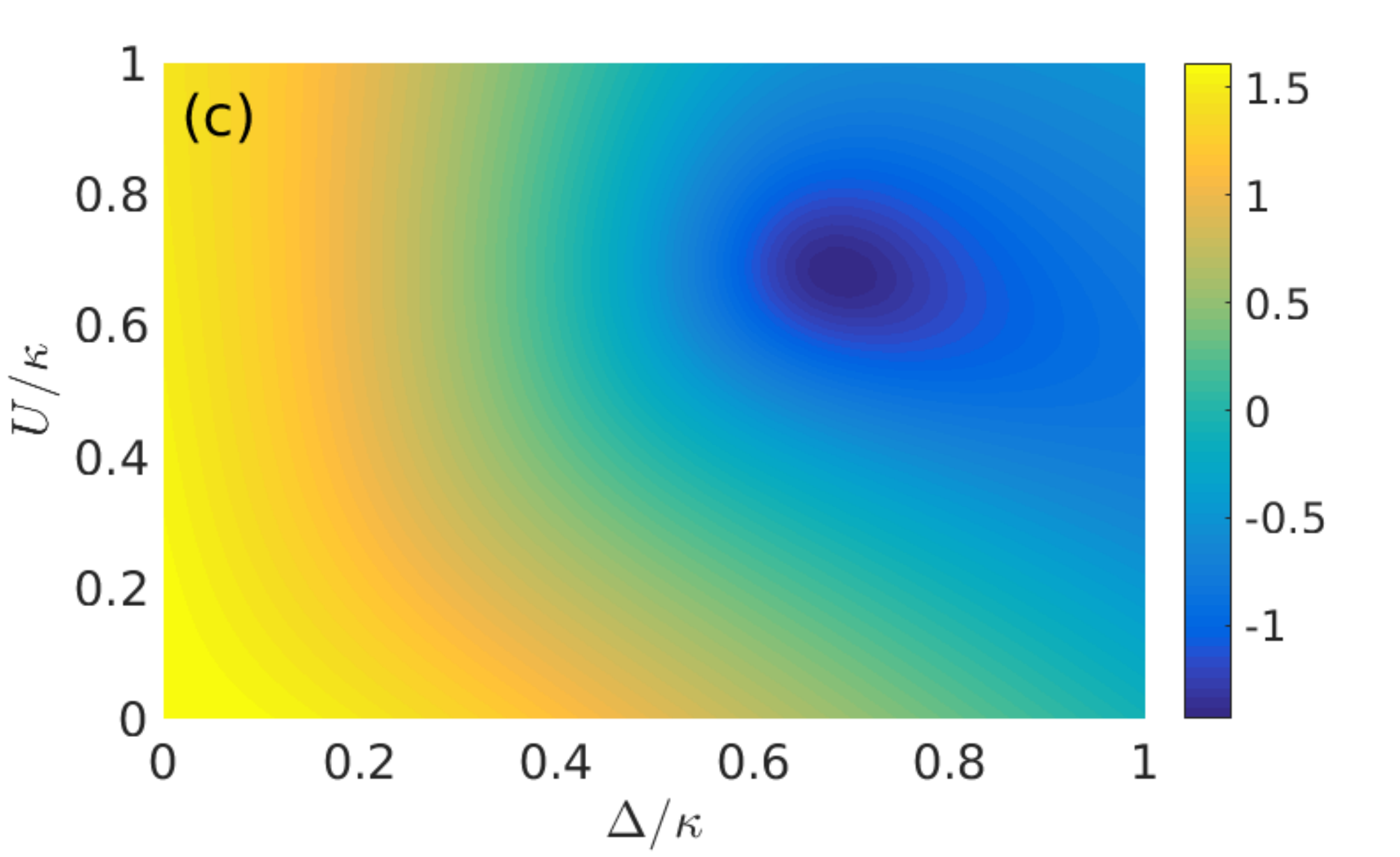}
	\includegraphics[width=0.45\linewidth]{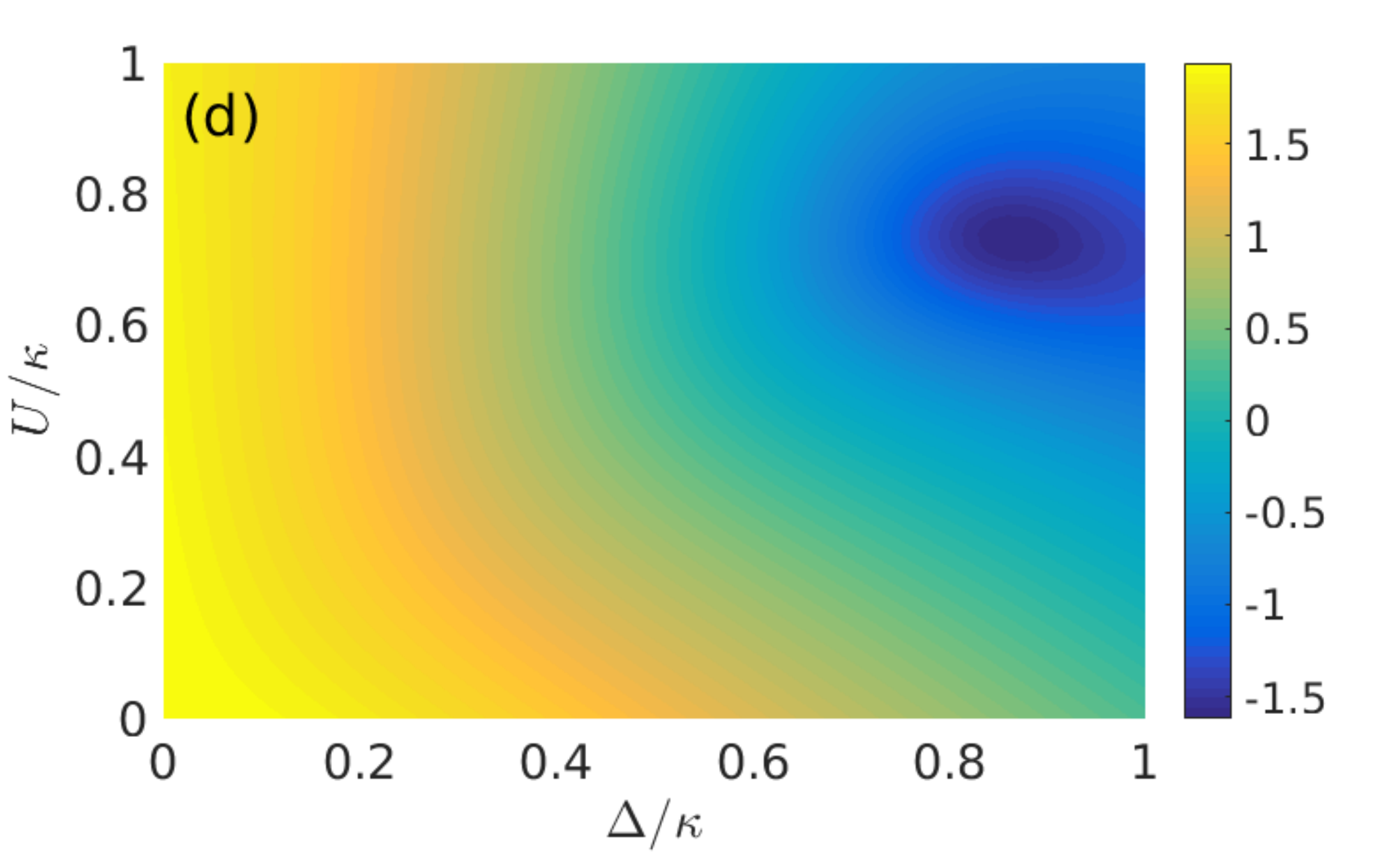}
	\caption{(Color online) Contour plots showing the variation of $\log_{10}{g_a^{(2)}(0)}$ as functions of $\Delta/\kappa$ and $U/\kappa$, for different values of $g/\kappa$ considered as: $g/\kappa = 1$ in (a), $g/\kappa = 1.5$ in (b), $g/\kappa = 2$ in (c), and $g/\kappa = 2.5$ in (d).}
	\label{upb1:fig3}
\end{figure*}

\noindent
In Fig.~\ref{upb1:fig2}(a), we show $g_a^{(2)}(0)$ as a function of $\Delta/\kappa$ with different moderate values of $g$. The values of $U$ is considered to be $U_\textrm{opt}$. For the values of $g$ considered in the plot, the optimal parameters from Eq.~\eqref{upb1:eq9} are obtained as, for $g/\kappa = 1$: $\Delta_\textrm{opt}/\kappa = 0.27$, $U_\textrm{opt}/\kappa = 0.98$; for $g/\kappa = 1.5$: $\Delta_\textrm{opt}/\kappa = 0.47$, $U_\textrm{opt}/\kappa = 0.71$; for $g/\kappa = 2$: $\Delta_\textrm{opt}/\kappa = 0.66$, $U_\textrm{opt}/\kappa = 0.69$; for $g/\kappa = 2.5$: $\Delta_\textrm{opt}/\kappa = 0.84$,
$U_\textrm{opt}/\kappa = 0.74$. It can be observed from Fig.~\ref{upb1:fig2}(a) that, as predicted from the optimal conditions calculated analytically, $g_a^{(2)}(0)$ shows a strong antibunching effect at the optimal values of $\Delta/\kappa$.

Fig.~\ref{upb1:fig2}(b), demonstrates the two-time second-order correlation function $g_a^{(2)}(\tau)$ which is calculated as:
\begin{align} \label{upb1:eq10} 
	g_a^{(2)}(\tau)= \frac{\langle a_1^\dagger (t) a_1^\dagger (t+\tau) a_1(t+\tau) a_1(t)\rangle}{\langle a_1^\dagger (t) a_1(t)\rangle^2}.
\end{align}	
This quantity, $g_a^{(2)}(\tau)$ is proportional to the joint probability of detecting one photon at time, $(t + \tau)$, provided another photon was detected at time, $t$, at that position \cite{knight2005introductory}. The plots show $g_a^{(2)}(\tau)$ under the optimal conditions for different values of $J$. We can observe that at $\tau=0$, $g_a^{(2)}(0) \approx 0$, and for other delay times $g_a^{(2)}(\tau)>g_a^{(2)}(0)$. Therefore, it clearly demonstrates that the emitted photons are antibunched and sub-Poissonian in nature. From Figs.~\ref{upb1:fig2}(a) and (b), one can observe that for the values of $U$ falling in the weak coupling regime, i.e.~$U < \kappa$, photon blockade can be realized owing to the quatum interference-inducing interaction, as verified by the optimal parameters. 

\begin{figure}[b]
	\centering
	\includegraphics[width=\linewidth]{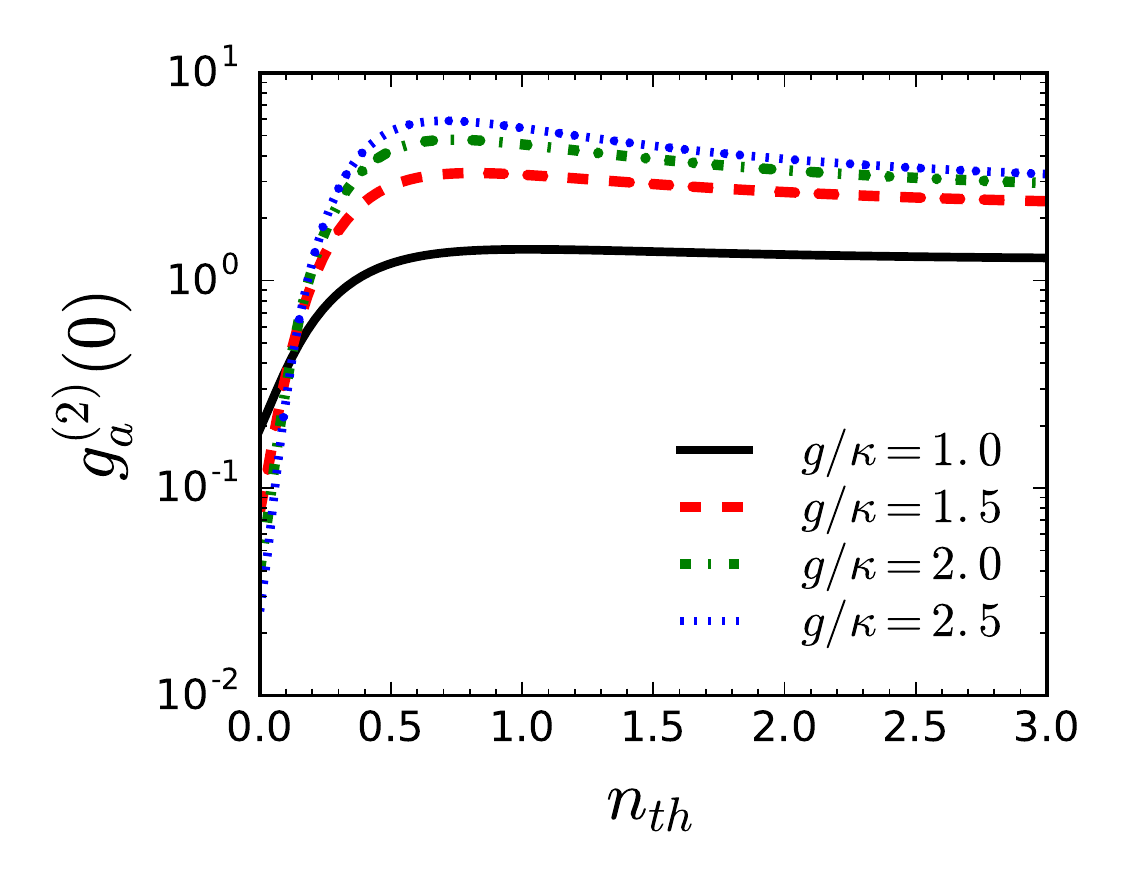}
	\caption{(Color online) Plot showing the effect of environmental temperature on photon blockade characteristics.}
	\label{upb1:fig4}
\end{figure}

In order to visualize the photon blockade effects more clearly, we show the contour plots of $g_a^{(2)}(0)$ in Fig.~\ref{upb1:fig3}, as functions of normalized detuning, $\Delta/\kappa$ and normalized nonlinear strength, $U/\kappa$. In Figs.~\ref{upb1:fig3}(a)-(d), the values of $g/\kappa$ are considered as: $g/\kappa = 1$ in (a), $g/\kappa = 1.5$ in (b), $g/\kappa = 2$ in (c), and $g/\kappa = 2.5$ in (d). The plots show that strong photon antibunching occurs exactly at the values predicted from the analytical calculations, in Eq.~\eqref{upb1:eq9}.

Next, we want to study the influence of environmental phonon population on the photon blockade characteristics. In Fig.~\ref{upb1:fig4}, we demonstrate $g_a^{(2)}(0)$ as a function of the bath phonon number, $n_{\rm{th}}$. For $g/\kappa=1$, $g_a^{(2)}(0)$ reaches $1$ at $n_{\rm{th}} \approx 0.5$, whereas, for $g/\kappa = 1.5$, $2$ and $2.5$, $g_a^{(2)}(0) \leq 1$ upto $n_{\rm{th}} \approx 0.25$. Therefore, it is evident that the environmental thermal phonon population has undesirable effect on the observation of photon blockade.

\begin{figure}
	\centering
	\includegraphics[trim={0 0 0 0},width=\linewidth]{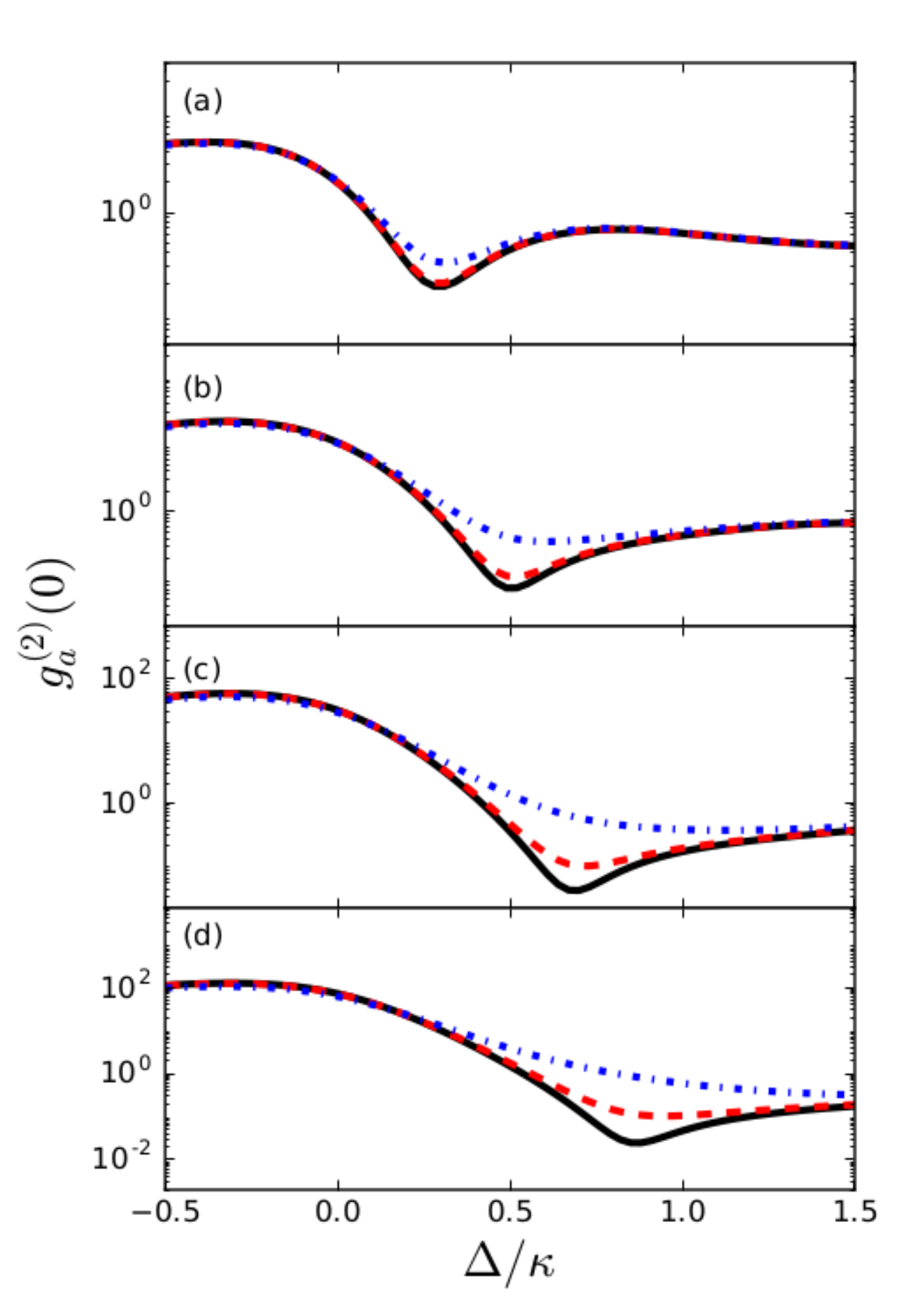}
	\caption{(Color online) Effect of pure dephasing. The black solid line represents $\gamma_p/\kappa = 0$, red dashed line is for $\gamma_p/\kappa = 0.001$ and the blue dash-dotted line denotes $\gamma_p/\kappa = 0.01$.}
	\label{upb1:fig5}
\end{figure}

Till now, in our analysis, we have not considered the effect of pure dephasing induced decoherences. Pure dephasing may arise from instability
of the laser drive, or coupling of the cavity modes to other mechanical modes, and due to this there can be perturbing effect on polarization , linewidth , transmittance, and photon statistics \cite{savona2013unconventional}. Therefore in the following, we analyze the effect of pure-dephasing on the antibunching properties of the cavity photons. The effects of pure dephasing can be modeled by adding another Lindblad term of the form $L_p(\rho) = \frac{\gamma_p}{2} \sum\limits_{j=1,2}{[2a_j^\dagger a_j \rho a_j^\dagger a_j - (a_j^\dagger a_j)^2 \rho - \rho (a_j^\dagger a_j)^2]}$, into the master equation, where $\gamma_p$ is the pure dephasing rate for the cavity modes. Figs.~\ref{upb1:fig5}(a)-(d) show the second-order correlation function $g_a^{(2)}(0)$ for different pure dephasing rates with different sets of optimized values. The values of $g$ is considered as: $g/\kappa = 1$ in (a), $g/\kappa = 1.5$ in (b), $g/\kappa = 2$ in (c), and $g/\kappa = 2.5$ in (d). The black solid line represents $\gamma_p/\kappa = 0$, red dashed line is for $\gamma_p/\kappa = 0.001$ and the blue dash-dotted line denotes $\gamma_p/\kappa = 0.01$. With increase in the pure dephasing rate, $g_a^{(2)}(0)$ increases near the optimal detuning. For higher values of pure-dephasing rates eg.~$\gamma_p = 0.01 \kappa$, $g_a^{(2)}(0)$ approaches classical Poissonian statistics.

\section{\label{sec:level1} CONCLUSION}
In conclusion, we analyzed the photon statistics in terms of the second-order correlation function, in a weakly driven optomechanical system, where two optical modes and one mechanical mode interact via a three-mode mixing. Due to this coupling, additional two-photon excitation pathways are created in the higher-frequency optical mode, which can be exploited to obtain the desired photon blockade characteristics in the system via quantum interference. We derived the optimal parameters required for strong photon blockade by solving the non-Hermitian Schr\"{o}dinger equation  containing the damping and decay in the system. The numerical calculations of the second-order correlation function obtained from solving the master equation show agreement with the analytical calculations. It is observed that even when the Kerr-type nonlinearity is weak, under the optimal conditions corresponding to the fulfillment of the quantum-interference effect, photon blockade is possible in the system.

\setcounter{secnumdepth}{0}
\section{ACKNOWLEDGMENTS}
B.~Sarma gratefully acknowledges a research fellowship from MHRD, Govt. of India.\\

\appendix
\section{APPENDIX: CALCULATION OF OPTIMAL PARAMETERS}
{
In steady-state, to solve the coupled equations, we follow the iterative method prescribed by Bamba et.~al \cite{bamba2011origin}, i.e. $C_{000}\gg \{C_{100}, C_{011}\}\gg \{C_{200}, C_{111}, C_{022}\}$. The optimal condition for complete photon blockade corresponds to when the probability of a photon in the state $|200\rangle$ equals zero. Therefore, at steady-state, the equations for $C_{100}$ and $C_{011}$ are given by: 

\begin{align} \label{upb1:eq11}
\nonumber	
\left(\Delta-i\frac{\kappa}{2}\right) C_{100}+gC_{011}+\Omega C_{000}=& 0,\\
\left(\Delta-i\frac{\kappa+\gamma}{2}\right) C_{011}+gC_{100}=& 0.
\end{align}  
These two equations give the coefficints $C_{100}$ and $C_{011}$ as
\begin{align} \label{upb1:eq12}
\nonumber
C_{100}=-\frac{\Omega (\Delta-i\frac{\kappa+\gamma}{2})}{(\Delta-i\frac{\kappa}{2})(\Delta-i\frac{\kappa+\gamma}{2})-g^2}C_{000},\\
C_{011}=\frac{\Omega g}{(\Delta-i\frac{\kappa}{2})(\Delta-i\frac{\kappa+\gamma}{2})-g^2}C_{000}.
\end{align}
Now, the equations for the other coefficients are reduced to
\begin{align} \label{upb1:eq13}
\nonumber
gC_{111}+\Omega C_{100}&=0,\\
\nonumber
\left[2\left(\Delta-i\frac{\kappa}{2}\right)-i\frac{\gamma}{2}\right] C_{111}+2 gC_{022}+\Omega C_{011}&=0,\\
2\left(\Delta + U -i\frac{\kappa+\gamma}{2}\right) C_{022}+2gC_{111}&=0.
\end{align}
The necessary and sufficient condition for the existence of nontrivial solutions of Eqs.~\eqref{upb1:eq13} gives the optimal parameters.
}

\end{document}